\documentclass[conference, a4paper, 10pt]{IEEEtran}

\usepackage{graphicx} 
\usepackage{amssymb} 
\usepackage{cite} 
\usepackage{enumerate} 
\usepackage{url} 
\usepackage[pdfborder={0 0 0}]{hyperref} 

\usepackage{color}

\usepackage[normalem]{ulem}

\usepackage{listings}
\usepackage{color}

\definecolor{dkgreen}{rgb}{0,0.6,0}
\definecolor{gray}{rgb}{0.5,0.5,0.5}
\definecolor{mauve}{rgb}{0.58,0,0.82}

\usepackage{soul}
\usepackage{tikz}
\usetikzlibrary{calc}

\makeatletter
\newif\if@anonymize

\@anonymizefalse 

\if@anonymize
\newcommand{\highlight@DoHighlight}{
	\fill [outer sep = -15pt, inner sep = 0pt, color=black]
	($(begin highlight)+(0,8pt)$) rectangle ($(end highlight)+(0,-3pt)$) ;
}

\newcommand{\highlight@BeginHighlight}{
	\coordinate (begin highlight) at (0,0) ;
}

\newcommand{\highlight@EndHighlight}{
	\coordinate (end highlight) at (0,0) ;
}

\newdimen\highlight@previous
\newdimen\highlight@current
\newlength{\item@width}

\DeclareRobustCommand*\anonymize{%
	\SOUL@setup
	\def\SOUL@preamble{%
		\begin{tikzpicture}[overlay, remember picture]
		\highlight@BeginHighlight
		\highlight@EndHighlight
		\end{tikzpicture}%
	}%
	\def\SOUL@postamble{%
		\begin{tikzpicture}[overlay, remember picture]
		\highlight@EndHighlight
		\highlight@DoHighlight
		\end{tikzpicture}%
	}%
	\def\SOUL@everyhyphen{%
		\discretionary{%
			\SOUL@setkern\SOUL@hyphkern
			\SOUL@sethyphenchar
			\tikz[overlay, remember picture] \highlight@EndHighlight ;%
		}{%
		}{%
			\SOUL@setkern\SOUL@charkern
		}%
	}%
	\def\SOUL@everyexhyphen##1{%
		\SOUL@setkern\SOUL@hyphkern
		\settowidth{\item@width}{##1}%
		\makebox[\item@width]{}%
		\discretionary{%
			\tikz[overlay, remember picture] \highlight@EndHighlight ;%
		}{%
		}{%
			\SOUL@setkern\SOUL@charkern
		}%
	}%
	\def\SOUL@everysyllable{%
		\begin{tikzpicture}[overlay, remember picture]
		\path let \p0 = (begin highlight), \p1 = (0,0) in \pgfextra
		\global\highlight@previous=\y0
		\global\highlight@current =\y1
		\endpgfextra (0,0) ;
		\ifdim\highlight@current < \highlight@previous
		\highlight@DoHighlight
		\highlight@BeginHighlight
		\fi
		\end{tikzpicture}%
		\settowidth{\item@width}{\the\SOUL@syllable}%
		\makebox[\item@width]{}%
		\tikz[overlay, remember picture] \highlight@EndHighlight ;%
	}%
	\SOUL@
}
\else
\newcommand{\anonymize}[1]{#1}
\fi
\makeatother

\title{Java Extensions for OMNeT++ 5.0}
\author
{\IEEEauthorblockN{Henning Puttnies, Peter Danielis, Christian Koch, Dirk Timmermann}
	\IEEEauthorblockA{University of Rostock\\
		Institute of Applied Microelectronics and Computer Engineering\\
		18051 Rostock, Germany\\
		Tel./Fax: +49 (381) 498-7277 / -1187251\\
		Email: henning.puttnies@uni-rostock.de}}

\begin{document}
	
	\maketitle
	
	\begin{abstract}
		On the one side, network simulation frameworks are important tools for research and development activities to evaluate novel approaches in a time- and cost-efficient way. On the other side, Java as a highly platform-independent programming language is ideally suited for rapid prototyping in heterogeneous scenarios. Consequently, Java simulation frameworks could be used to firstly perform functional verification of new approaches (and protocols) in a simulation environment and afterwards, to evaluate these approaches in real testbeds using prototype Java implementations. Finally, the simulation models can be refined using real world measurement data. Unfortunately, there is to the best of our knowledge no satisfying Java framework for network simulation, as the OMNeT++ Java support ended with OMNeT++ version 4.6. Hence, our contributions are as follows: we present Java extensions for OMNeT++ 5.0 that enable the execution of Java simulation models and give a detailed explanation of the working principles of the OMNeT++ Java extensions that are based on Java Native Interface. We conduct several case studies to evaluate the concept of Java extensions for OMNeT++. Most importantly, we show that the combined use of Java simulation models and C++ models (e.g., from the INET framework) is possible.
	\end{abstract}

	\section{Introduction}
	Network simulators are perfectly suitable for the early evaluation of innovative approaches (e.g., novel applications or protocols for communication, control, or security). Popular simulators like OMNeT++ are recommendable as they offer many existing modules suitable for reuse, generally have a good usability, and are seriously tested. Furthermore, the Java programming language is perfectly suitable for rapid prototyping as it is very predictable, easy to debug, and highly platform independent. As a consequence, Java programs can be deployed on different platforms with minimal adaptation effort, which enables the evaluation of an approach on many different devices in heterogeneous IoT scenarios.
	
	Therefore, it is an interesting approach to combine OMNeT++ and Java like it was possible using the Java extensions for the OMNeT++ Versions 3.X to 4.6. Firstly, using Java and OMNeT++ enables to evaluate an approach using simulation models written in Java. Secondly, the Java simulation models can be used to develop a platform independent Java prototype implementation and thus, quickly evaluate research approaches in real world scenarios. As a result it is possible to feed the results (e.g., realistic computation times) back into the simulation models. Our contributions are as follows:
	\begin{itemize}
		\item We generated and present Java extensions for OMNeT++ 5.0 derived from the existing Java extensions for OMNeT++ 4.6 and give a brief overview of the Java capabilities of other existing simulation frameworks.
		\item We explain in detail how the Java extensions work in combination with OMNeT++. In contrast to the existing documentation of the Java extensions that focus on how to use the Java extensions, we turn our attention to their functional principles. We want to help other researchers in understanding the Java extensions for OMNeT++ and encourage them to use Java simulation models in OMNeT++. 
		\item We conduct several case studies and evaluate the effort of porting a real Java application to a Java simulation model. Moreover, we present the possibility to combine Java simulation models in OMNeT++ with existing C++ models (e.g., the INET library). This is an important case study, as the reuse of existing modules can tremendously reduce the time to develop a new simulation model. To the best of our knowledge, this is the first analysis and proof of concept implementation of the combination of Java simulation modules and existing C++ modules (e.g., INET framework), besides the use of the OMNeT++ simulation kernel. 
		\item The entire system including all source code is freely available for download. We share a virtual machine (VM) for VirtualBox \cite{.az} running Ubuntu. Therefore, the system is running "out of the box" without any platform dependencies (e.g., compilers, Java Virtual Machine) or the need for configuration (e.g., paths, environment). This is an important point as the OMNeT++ Java extensions base on Java Native Interface (JNI), which highly depends on the environment (operating system, Java virtual machine, paths etc.). Thus, sharing a VM facilitates the use of the Java extensions.
	\end{itemize}

	\section{Related Work}
	As the generation of Java extensions (interfaces) for a C++ simulation framework (like OMNeT++) requires substantial efforts, we analyze the Java capabilities of existing simulators in the following. 
	
	The Network Simulator 3 (NS-3) \cite{Henderson.2008,.ar} is a very popular simulation framework for computer networks and the successor of NS-2, although NS-3 was developed from scratch. It is written in C++ and to the best of our knowledge, there is at the time of writing no existing approach to integrate Java simulation models into NS-3. 
	
	Java Network Simulator (JNS) \cite{.av} is a Java implementation of NS-2. Nevertheless, the development stopped in 2010 and it always had less features than NS-2 as stated by the developers.
	
	JNetworkSim \cite{.aw} is a modular and fast simulator developed at Stanford University. Nevertheless, it focuses on the simulation of network switches rather than entire network topologies. Furthermore, it was lastly updated in 2007.
	
	Another Java simulation framework is the Probabilistic Wireless Network Simulator (JProwler) \cite{.ax} that is a simple wireless network simulator. It is (as stated by the developers) small and lightweight. However, as the download website has not been updated since 2008, we assume that it is no longer maintained.
	
	Psimulator2 \cite{.bb} is a basic graphical network simulator originally developed at the Czech Technical University in Prague. Although it is still under maintenance, the purpose of this simulation framework is the teaching of basic networking topics. In contrast, we focus on research activities as motivated in the introduction.
	
	Furthermore, there is a free network simulator written in Java \cite{.at}, which has the ability to reconfigure routing and topology. It is stated to be cycle-based and to model flit-level communication. Assumingly, it is out of maintenance as the initial author James Hanlon finished his Ph.D. in 2014.
	
	After analyzing related works, we can conclude that there is no existing simulation framework combining a popular simulator (good for reuse) and the ability to development simulation models in Java (good for rapid prototyping). We address this need with the Java extensions for OMNeT++ 5.0.

	\section{Concept of Java Extensions for OMNeT++}
	
	\subsection{Working Principles of Java Extensions for OMNeT++}
\begin{figure}[htbp]
		\centering
		\includegraphics[width=0.45\textwidth]{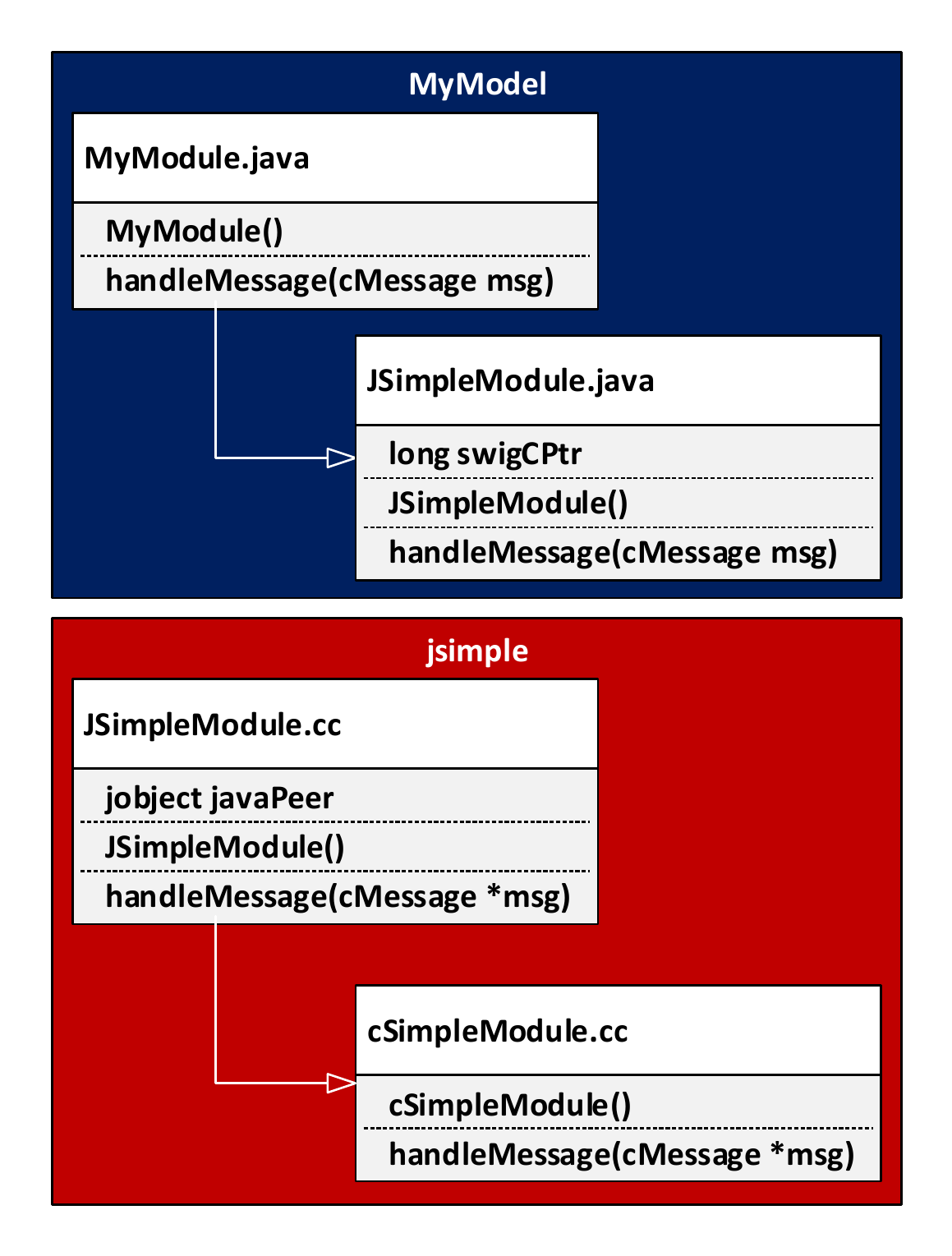}
		\caption{Class diagrams of extension modules (e.g., JSimpleModule) that form the Java extensions. Blue: Java code, Red: C++ code. \textit{JSimpleModule.java} is a wrapper for \textit{JSimpleModule.cc}. Therefore, it has the member \textit{long swigCPtr} that is a pointer to \textit{JSimpleModule.cc}. Vice versa, \textit{JSimpleModule.cc} has the member \textit{jobject javaPeer} that is a pointer to \textit{JSimpleModule.java}.}
		\label{fig:CD}
	\end{figure}

	The Java extensions for OMNeT++ base on the JNI \cite{.au}. To allow the use of Java simulation models, a new simulation executable is generated. This simulation executable (jsimple) consists of the entire OMNeT++ simulation kernel as well as several extension modules. As depicted in the class diagrams (see Fig.\ref{fig:CD}), the developer can write a Java simulation model (e.g., \textit{MyModule.java}) that inherits from \textit{JSimpleModule.java}, which is a Java wrapper for the C++ class \textit{JSimpleModule.cc}. \textit{JSimpleModule.cc} is an extension class for OMNeT++ that implements the interface between the Java simulation modules (e.g., \textit{MyModule.java}) and the OMNeT++ simulation kernel (via \textit{cSimpleModule.cc}).
	
\begin{figure}[htbp]
\centering
\includegraphics[width=0.45\textwidth]{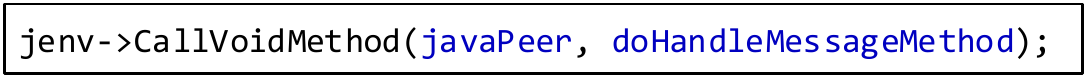}
\caption{Calling a Java method from C++ (\textit{JSimpleModule::handleMessage()}): jenv = pointer to java environment; CallVoidMethod = calls a method of an object; javaPeer = pointer to the java peer of this object (in this case: \textit{JSimpleModule.java}); doHandleMessageMethod = method ID of \textit{handleMessage()} }
\label{fig:call_cpp}
\end{figure}

\begin{figure}[htbp]
\centering
\includegraphics[width=0.45\textwidth]{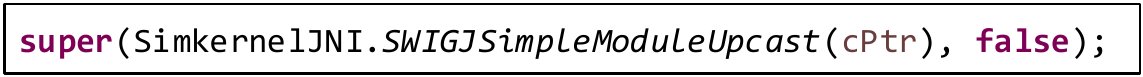}
\caption{Calling a C++ method from Java (constructor of \textit{JSimpleModule.java}): super = constructor of ancestor (\textit{cSimpleModule}); SimkernelJNI = class holding Java wrappers for all native (C++) methods; cPtr = pointer to corresponding C++ class (\textit{JSimpleModule.cc}); false = dummy value}
\label{fig:call_java}
\end{figure}

\begin{figure}[htbp]
\centering
\includegraphics[width=0.45\textwidth]{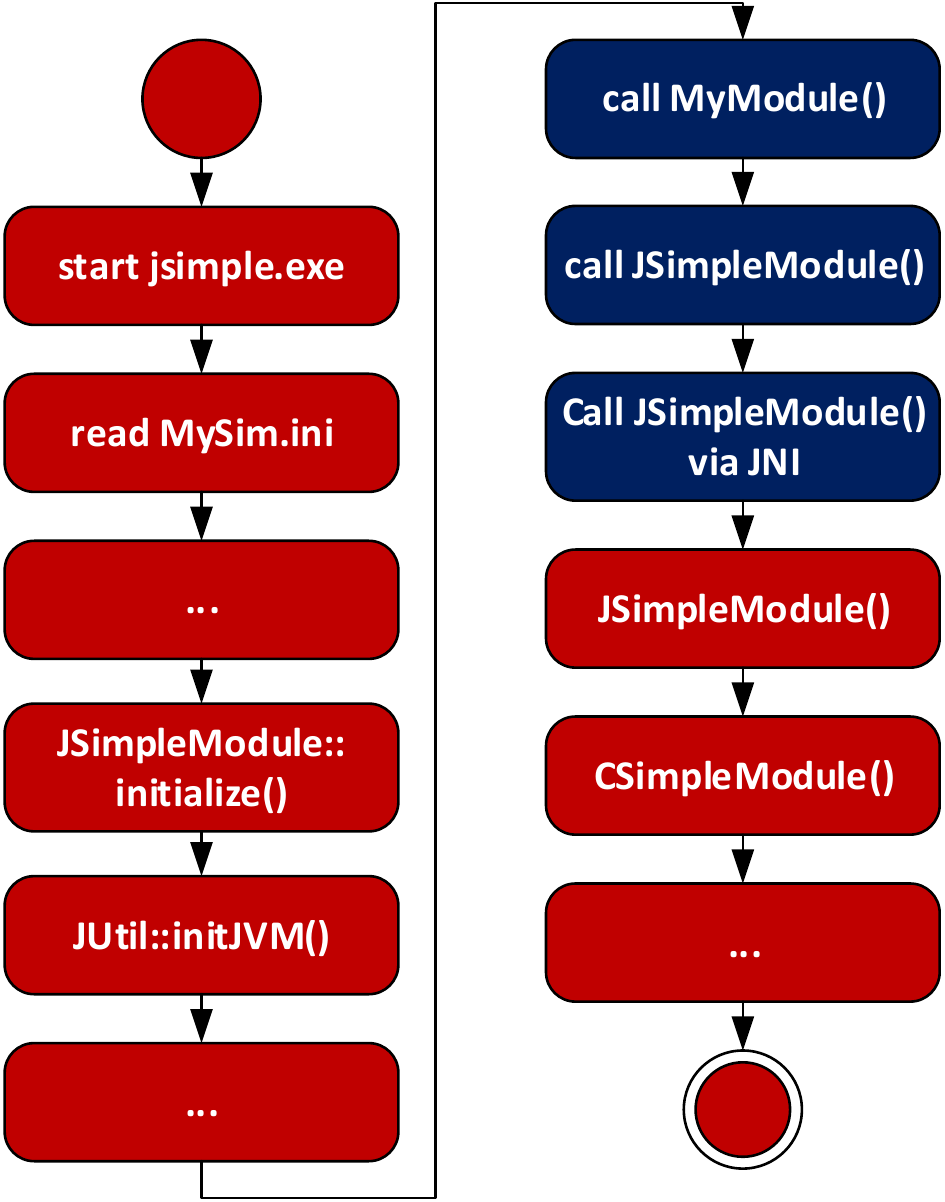}
\caption{Flow chart of the OMNeT++ with Java extensions executing a Java simulation model. Blue: Java code, Red: C++ code}
\label{fig:FC}
\end{figure}

	Fig. \ref{fig:FC} depicts the flow chart of an exemplary execution of OMNeT++ with Java extensions. The program \textit{jsimple.exe} is started as simulation executable and reads the corresponding \textit{*.ini} file (e.g. \textit{MySim.ini)}. Let us assume, that \textit{MySim.ini} loads a \textit{*.ned} file that uses \textit{MyModel} (a Java simulation model) from Fig.\ref{fig:CD}. The \textit{JSimpleModule::initialize()} method is calling the \textit{JUtil::initJVM()} method to start the Java virtual machine (JVM). As the JVM is a shared library that can execute Java bytecode in \textit{*.class} files, it is possible to execute Java simulation models. In the Java simulation model, the constructor of \textit{MyModule} is called that in turn calls the constructor of \textit{JSimpleModule} and C++ code via JNI. It is possible to call a Java method from C++ code and vice versa. Figures \ref{fig:call_cpp} and \ref{fig:call_java} depict and explain exemplary code snippets of the JNI calls. 
	
	The \textit{SimkernelJNI\_registerNatives()} method is used in original Java Extensions and the Java Extensions for OMNeT++ 5.0. A PERL script provided by the OMNeT++ 4.6 (\textit{registernatives.pl}) automatically generates this method. \textit{SimkernelJNI\_registerNatives()} registers all C++ methods that are wrapped and accessible from Java code. The purpose of the \textit{SimkernelJNI\_registerNatives()} method is to check the signature of every C++ method that is wrapped and accessible from Java code, before this method is actually used \cite{.as}. This substantially increases the stability of the system, as there are more than 1000 methods accessible from Java code and otherwise the signatures are only checked at runtime, which might lead to unstable behaviour.
	
	\subsection{Generating the Java Extensions}
	In what follows, we will describe the process of generating the Java extensions. The same approach might be used to generate wrappers for any C++ simulation models to reuse them in conjunction with Java simulation models. The Java wrapper classes for OMNeT++ are automatically generated using SWIG \cite{Beazley.1996,.bc}, which is a well-documented and powerful tool that can be used to generate an interface between ANSI C/C++ and many other high level languages (e.g., Java, Chicken, and Python). SWIG uses the JNI API for this interface. Files that are of the type \textit{*.i} (interface files) are used to configure the process of generating the interface correctly. There are several specifics when using SWIG and generating Java Extensions for C++ code (e.g., OMNeT++). Therefore, we focus on the most important ones in the following.
	
	The overloading of operators is available in C++ but not in Java. As a consequence, these operators are wrapped into methods (e.g. "=" is wrapped into \textit{set()}, "==" into \textit{sameAs()}, and "++" into \textit{incr()}). This can be done automatically by applying the \textit{\%rename} directive in the SWIG interface file.
	
	As there are no pointers available in Java, specific SWIG pointers exist, which are wrappers holding the address of the corresponding C++ object as a Java long.
	
	SWIG can handle namespaces but ignores them in the names of the Java wrapper code. Therefore, a method \textit{Foo::Bar()} in C++ is wrapped into a method \textit{Bar()} in Java by SWIG. If two methods ,which have the same name, are defined in two separate namespaces (e.g., \textit{Foo1::Bar()} and \textit{Foo2::Bar()}), SWIG maps them to the same Java method name (e.g., \textit{Bar()}) that is therefore defined multiple times and the compile process crashes. The solution is to use the \textit{\%rename} directive to rename \textit{Foo1::Bar()} into \textit{Foo1\_Bar()}.
	
	SWIG does not support (and hence ignores) nested classes (class definitions in classes). The solution consists in a redefinition of the inner class in the SWIG interface file (*.i). This redefinition makes the inner class globally visible and hence SWIG generates a wrapper for this class.
	
	\subsection{Differences between the Java Extensions for OMNeT++ 4.6 and OMNeT++ 5.0}
	As the reuse of existing software is a powerful method to speed up the development process and improve its results, we have not developed the Java extensions for OMNeT++ 5.0 from scratch. In constrast, our Java extensions are derived from the Java extensions for OMNeT++ 4.6. In the following, we describe the difference between the Java extensions for OMNeT++ 4.6 and the Java extensions for OMNeT++ 5.0. 
	
	Whereas there are only small differences between the Java extensions for the OMNeT++ versions between 4.1 and 4.6 (two additional Lines for cCompoundModule in the SWIG interface file), we apply multiple changes:
	
	In the SWIG interface file, we have to include all OMNeT++ headers explicitly (e.g., \textit{simkerneldefs.h} $\rightarrow$ \textit{omnetpp\textbackslash simkerneldefs.h}), as SWIG does not “follow” C++ includes. As described previously, SWIG is able to handle namespaces. Consequently, we have to refer to all C++ classes with the namespace prefix (\textit{omnetpp::}) in the SWIG interface file. All generated Java wrapper classes are referred to without a prefix. Furthermore, we removed several header files (e.g., \textit{cevent.h} and \textit{ceventheap.h}) from the SWIG interface file and added a few (e.g., \textit{cmessageheap.h}).
	
	In the extension classes (JSimpleModule, JMessage, JUtil), we moved all declarations and definitions into the namespace omnetpp to cope with this namespace introduced by OMNeT++ 5.0. This allows a dynamic name solving within the source code of the C++ classes (e.g., \textit{JSimpleModule.cc}) similar to the Java extensions for OMNeT++ 4.6.
	
	The \textit{registernatives()} method now registers a total number of 1926 Java wrapper methods. In comparison, 1400 Java wrapper methods where available in the Java extensions for OMNeT++ 4.6.

	\section{Case Studies: Java Simulation Models in OMNeT++ 5.0}
	
	Firstly, we used the Jsamples project that is provided with the Java Extensions for OMNeT++ 4.6. The Jsamples project consists of several sample applications (TicToc etc.) that serve as tutorial similarly to their C++ counterparts. We used this project to test our Java extensions for OMNeT++ 5.0 successfully.
	
	\subsection{Converting a Java UDP Ping Implementation to a Simulation model}
	As first cast study, we developed a UDP-based Ping implementation in Java to evaluate the effort for porting a real Java application to a Java simulation model. Firstly, we tested the application in a real network consisting of two Galileo uno boards \cite{.ad}. Secondly, we converted this real Java application to a Java simulation model that is executable using OMNeT++ 5.0 with Java extensions. For the functionality of the Java application, the porting process turned out to be easy. However, regarding the API for the communication interface (in this case: UDP sockets), the Java code has to be adapted to the working principles of OMNeT++ (especially communication based on \textit{cMessages}). This is a noticeable effort as the API of system calls (e.g., UDP sockets) and the API of OMNeT++ (or INET) are completely different.

	\subsection{Using the INET Framework in Conjunction with Java Simulation Models}
\begin{figure}[htbp]
	\centering
	\includegraphics[width=1.0\columnwidth]{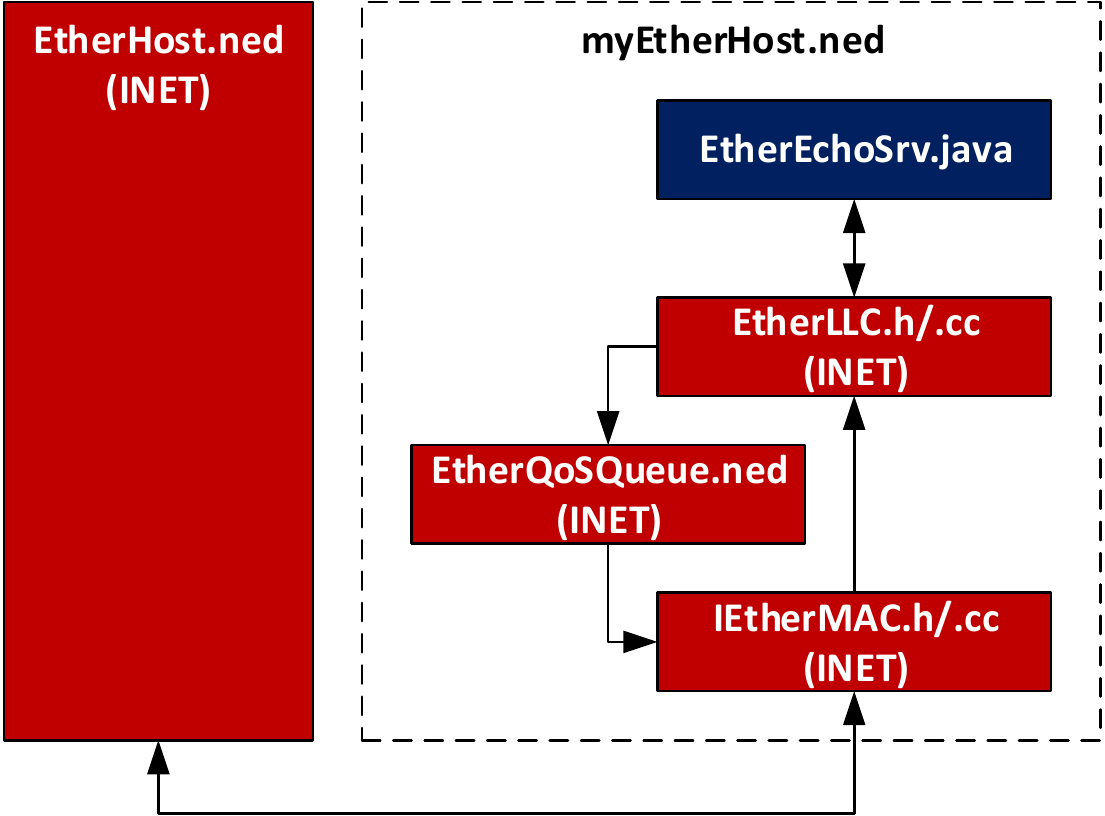}
	\caption{Overview of a simulation model that consists of both, Java simulation modules and existing C++ modules (here: from the INET framework). Blue: Java code, Red: INET code (C++ and \textit{*.ned}). }
	\label{fig:JI}
\end{figure}

	As second case study, we evaluated the interface between the modules of the INET framework and Java simulation models. This is an important case study as the reuse of existing INET modules is an outstanding way to speed up the development of new simulation models. As first step, we executed the INET wireless tutorial that entirely consists of C++ modules using the jsimple simulation executable (simulation kernel and Java extensions). Hereby, we showed that the INET modules can be used in \textit{*.ned} files and executed by jsimple. As second step, we developed a simulation model consisting of both Java and C++ modules to analyze whether they can be connected and communicate with each other. Our simulation model (see Fig.\ref{fig:JI}) is derived from the INET example \textit{"inet/examples/ethernet/lans/twoHosts.ini"}. We used the \textit{EtherHost} from INET and connected it with our own implementation called \textit{myEtherHost}. The module \textit{myEtherHost} consists of \textit{EtherLLC}, \textit{EtherQoSQueue}, and \textit{IEtherMAC} from INET written in C++ and \textit{EtherEchoSrv}, which is a Java module. The evaluation showed, that \textit{EtherEchoSrv} (Java) can correctly register to \textit{EtherLLC} (C++) using the \textit{Ieee802Ctrl} (C++) control structure form INET. Moreover, the module \textit{EtherEchoSrv} (Java) correctly receives Ethernet packets from the INET module and sends back valid Ethernet packets, which are accepted by the \textit{EtherLLC} and \textit{IEtherMAC} (both C++) modules from INET. Thus, we showed that even the combination of INET modules and Java simulation modules is possible. 
	
	\subsection{Limitations of the OMNeT++ Java Extensions}
	Although the porting of a simulation model to a real application is straight forward for simple programs, there are several limitations of OMNeT++ and the Java extensions.

	Multithreading is not supported within one application. Therefore, extensive implementations using multiple threads have to be reduced into one thread or partitioned into several LPs (logical processes). This however is a general disadvantage of OMNeT++ \cite{.ay} and thus does not only apply to the Java extensions.

	Moreover, the modelling of concurrent behaviour using while loops is not adequate, because the simulation would freeze if an operation is not atomic. It is impossible to send a packet and receive the response within one execution of the \textit{handleMessage()} method, while a similar behaviour is possible within the \textit{main()} method of a Java application. This might necessitate manual modifications of the Java code. Nevertheless, this is not a drawback introduced by the Java extensions.

	\section{Summary and Outlook}
	In this paper, we presented the Java extensions for OMNeT++ 5.0 and describe the generation of the OMNeT++ Java extensions as well as their functional principles. Furthermore, we conducted several case studies to evaluate the effort of porting real Java applications to Java simulation modules. Although our Java extensions are derived from the existing Java extensions for OMNeT++ 4.6, we enable the possibility to interface Java simulation modules and C++ modules (e.g. from the INET library). This was not examined before but is important for the reuse of existing modules and therefore for the utility of the Java extensions.

	The generation of Java extensions for OMNeT++ 5.1 would be of interest for future work. This should be easier than the generation of the extensions for OMNeT++ 5.0 as the difference between OMNeT++ 5.0 and OMNeT++ 5.1 is much smaller than the changes between OMNeT++ 4.6 and OMNeT++ 5.0.

	\section{Downloading the Source code}
	The entire system (OMNeT++ 5.0, Java extensions, and case studies) is available online\footnote{\href{https://bwsyncandshare.kit.edu/dl/fi8R6skmuBPh6UfXHWzcgBxt/.zip}{https://bwsyncandshare.kit.edu/dl/fi8R6skmuBPh6UfXHWzcgBxt/.zip}} as an Ubuntu virtual machine for VirtualBox. In contrast to all previous versions of the Java extensions for OMNeT++, there is neither a need to regenerate the Java extensions nor to recompile the jSimple simulation executable (OMNeT++ simulation kernel with Java extensions). This is an enormous benefit as JNI-based systems are highly dependent on platforms and paths.
	
	\bibliographystyle{IEEEtran}
	

\end{document}